\documentclass[letterpaper]{article} % DO NOT CHANGE THIS
\usepackage{aaai25}  % DO NOT CHANGE THIS
\usepackage{times}  % DO NOT CHANGE THIS
\usepackage{helvet}  % DO NOT CHANGE THIS
\usepackage{courier}  % DO NOT CHANGE THIS
\usepackage[hyphens]{url}  % DO NOT CHANGE THIS

\usepackage{times}  % DO NOT CHANGE THIS
\usepackage{helvet}  % DO NOT CHANGE THIS
\usepackage{courier}  % DO NOT CHANGE THIS
\usepackage[hyphens]{url}  % DO NOT CHANGE THIS
\usepackage{graphicx} % DO NOT CHANGE THIS
\usepackage{xcolor}
\definecolor{lgray}{rgb}{0.9, 0.9, 0.9}
\usepackage[utf8]{inputenc}
\DeclareUnicodeCharacter{00E9}{\'{e}}

\usepackage{tabularx}
\usepackage{multirow}
\usepackage{amsmath}
\usepackage{amssymb}
\usepackage{color}
\usepackage{graphicx}
\usepackage{xcolor}
\usepackage{colortbl}
\usepackage{tikz}
\usepackage{amsmath}
\usepackage{subfigure}
\usepackage{booktabs}
\usepackage{nameref}
\usepackage{inconsolata}

\usepackage{tcolorbox}

\usepackage[ruled,linesnumbered,algo2e]{algorithm2e}
\usepackage{url}

\usepackage{graphicx} % DO NOT CHANGE THIS
\usepackage{natbib}  % DO NOT CHANGE THIS AND DO NOT ADD ANY OPTIONS TO IT
\usepackage{caption} % DO NOT CHANGE THIS AND DO NOT ADD ANY OPTIONS TO IT
\usepackage{algorithm}
\usepackage{algorithmic}

\usepackage{newfloat}
\usepackage{listings}
\DeclareCaptionStyle{ruled}{labelfont=normalfont,labelsep=colon,strut=off} % DO NOT CHANGE THIS
\floatstyle{ruled}
\newfloat{listing}{tb}{lst}{}
\floatname{listing}{Listing}
\title{PhishAgent: A Robust Multimodal Agent for Phishing Webpage Detection}
\author{
{ Tri Cao\textsuperscript{1 }\equalcontrib\thanks{Corresponding Author: Tri Cao (\tt{tricao@nus.edu.sg}).}, 
Chengyu Huang\textsuperscript{1}\equalcontrib, 
Yuexin Li\textsuperscript{1}\equalcontrib, 
Huilin Wang\textsuperscript{1}, Amy He\textsuperscript{3}},\\
{ Nay Oo\textsuperscript{2}, 
Bryan Hooi\textsuperscript{1}}}
 \affiliations{
\textsuperscript{\rm 1}National University of Singapore, \\\textsuperscript{\rm 2}NCS Cyber Special Ops-R\&D, \\\textsuperscript{\rm 3}Massachusetts Institute of Technology\\}
\begin{document}
\maketitle

\begin{abstract}
Phishing attacks are a major threat to online security, exploiting user vulnerabilities to steal sensitive information. Various methods have been developed to counteract phishing, each with varying levels of accuracy, but they also face notable limitations. In this study, we introduce PhishAgent, a multimodal agent that combines a wide range of tools, integrating both online and offline knowledge bases with Multimodal Large Language Models (MLLMs). This combination leads to broader brand coverage, which enhances brand recognition and recall. Furthermore, we propose a multimodal information retrieval framework designed to extract the relevant top $k$ items from offline knowledge bases, using available information from a webpage, including logos and HTML. Our empirical results, based on three real-world datasets, demonstrate that the proposed framework significantly enhances detection accuracy and reduces both false positives and false negatives, while maintaining model efficiency. Additionally, PhishAgent shows strong resilience against various types of adversarial attacks.
\end{abstract}

\section{Introduction}
% What is agent, the benefit of the agent, the limitation of agent, the limitation of other approach
Phishing attacks pose a serious threat to online security, as cybercriminals continuously improve their methods to trick users into disclosing sensitive information by pretending to be legitimate entities. According to the Anti-Phishing Working Group (APWG), there were 1,077,501 reported phishing attacks in the fourth quarter of 2023, contributing to nearly five million attacks throughout the year—the highest number ever recorded \cite{APWG2023}. These deceptive sites result in substantial financial losses, as the FBI reported that U.S. businesses suffered losses exceeding \$12.5 billion due to phishing in 2023, up from \$10.3 billion in 2022 \cite{FBI2023}. 
These numbers emphasize the urgent necessity for robust automated phishing detection methods and the pressing need to confront this escalating threat.

 Many approaches have been developed to combat phishing, each achieving varying degrees of accuracy but also facing significant limitations. Conventional approaches, such as heuristic-based methods \cite{ urlnet, urltran, url_1, hinphish, cantina+, stackmodel, d-fence}, blacklists \cite{google_safe_browsing, openphish, phishtank}, and rule-based systems \cite{afroz2011phishzoo}, have been effective to some extent in analyzing webpage characteristics, maintaining lists of known phishing URLs, and applying predefined rules based on known phishing patterns. However, these methods are often static and not up-to-date, relying on fixed criteria that do not adapt to new phishing techniques. For example, a heuristic method may only detect URLs that fit predefined patterns, missing new variations. This can lead to delays in detecting evolving phishing threats until the rules are manually updated. Reference-based approaches compare suspected phishing sites against a knowledge base of legitimate webpages for various brands \cite{visualphishnet, phishpedia, phishintention, dynaphish, li2024knowphish}, achieving good results but facing challenges in maintaining a comprehensive and current knowledge base. Search engine-based methods generate query strings from webpage content and analyze search results \cite{ zhang2007cantina, huh2012phishing, jain2018two, varshney2016phish, chang2013phishing, chiew2015utilisation, rao2019jail}, showing promise but being prone to false positives and sensitive to changes in search engine algorithms, which can affect their effectiveness. Recently, Multimodal Large Language Model (MLLM)-based approaches have demonstrated high accuracy in detecting phishing webpages by leveraging advanced text and image processing capabilities \cite{koide2023detecting}. Nevertheless, MLLMs are susceptible to adversarial attacks \cite{li2024knowphish, cui2024robustness} and may struggle with local brands, as information about these brands is limited and unfamiliar to the MLLMs, leading to potentially erroneous decisions.
 % While these methods have yielded positive results, they still face limitations in performance due to limited knowledge bases, and vulnerability to attacks.
 % However, these methods can be static and non-interactive, often struggling to keep up with evolving phishing techniques and experiencing delays in detecting new threats.
 
In light of these challenges, autonomous agents offer a promising solution. Defined as Large Language Models (LLMs) interacting with a set of tools, these agents are particularly suited for solving complex tasks due to their ability to analyze information from the tools and make decisions. This capability has been successfully demonstrated in many tasks \cite{park2023generative, shen2024hugginggpt, chen2024agent, zeng2023agenttuning}. Recently, GEPAgent \cite{wang2024automatedphishingdetectionusing}, an agent-based phishing detection approach, was introduced, demonstrating high accuracy in detecting phishing webpages through interactions between MLLMs and an online knowledge base across multiple iterations. However, it also exhibited a significant limitation in terms of execution time, taking over 10 seconds per sample. 

To solve these challenges, we propose PhishAgent, a multimodal agent specifically designed for phishing webpage detection with low latency. PhishAgent integrates a comprehensive set of tools, combining both online and offline knowledge bases. GEPAgent \cite{wang2024automatedphishingdetectionusing} require multiple iterations of interaction between the agent and knowledge bases to refine results, leading to high latency. In contrast, PhishAgent is designed to use only one such iteration, yet still achieves high performance, thus significantly reducing detection time. Its various modules are interconnected through the Agent Core, which functions as the central module, integrating all the other modules and making decisions. Particularly, the offline knowledge base can cover local brands that are not indexed by search engines, while the online knowledge base can cover very new webpages that the offline knowledge may not include. This combination results in wider brand coverage, thereby increasing brand recognition and recall. Moreover, MLLMs in PhishAgent make decisions based on the information queried from the knowledge base rather than solely based on its internal knowledge, which enhances result reliability.

In addition, we make improvements to several auxiliary tools to enhance their accuracy. Specifically, recognizing the limitations of querying the most related webpage in the offline knowledge base based solely on exact matching between the extracted brand name of the input webpage and the brand names existing in the knowledge base through the textual modality \cite{li2024knowphish}, we introduce a multimodal information retrieval framework. This framework enhances the analysis of the input webpage by leveraging all available information from the webpage, such as logos and HTML, to retrieve the top $k$ relevant items from the offline knowledge base. These relevant items are used in the subsequent steps of our pipeline to ultimately determine whether the webpage is a phishing webpage or not.

In summary, our work makes three
main contributions:
\begin{itemize}
    \item \textbf{PhishAgent}: A multi-modal agent tailored for low-latency phishing webpage detection, combining both online and offline knowledge bases with MLLMs and various useful tools.  
    \item \textbf{Multi-modal Retriever}: A multimodal module which retrieves the top $k$ brands from an offline knowledge base, utilizing all available information from the webpage, such as logos and HTML.
    \item \textbf{Empirical Results}: Our empirical results on three real-world datasets show that the proposed framework notably enhances detection accuracy while preserving model efficiency. 
    Additionally, PhishAgent also demonstrates robustness against various types of adversarial attacks. 
\end{itemize}

\section{Related Works}

\paragraph{Conventional Approaches} These include methods relying on heuristics, blacklists, and rule-based systems. Heuristic methods analyze webpage characteristics such as URL, HTML structure, and suspicious keywords \cite{garera2007framework, sheng2010falls, zhang2007cantina, urlnet, urltran, url_1, hinphish, cantina+, stackmodel, html_url_1, d-fence}. Blacklists check incoming URLs against known phishing lists \cite{google_safe_browsing, openphish, phishtank}, while rule-based systems use predefined rules based on known phishing patterns \cite{afroz2011phishzoo}. However, these methods tend to be static, depending on fixed criteria that fail to adjust to new phishing techniques. For instance, a heuristic approach might only identify URLs that match established patterns, overlooking new variations. This can result in delays in identifying emerging phishing threats until the rules are manually revised.

\paragraph{Reference-based Approaches} These compare target webpage information to a known set of brand information. They create a brand knowledge base (BKB) containing logos, aliases, and legitimate domains, and a detector backbone that uses this BKB for detection \cite{li2024knowphish, dynaphish}. To determine whether a webpage is phishing or legitimate, these systems first identify the target brand of the webpage. If the webpage is found to have the intent of a particular brand but its domain does not align with the brand's authentic domains, it is classified as phishing. However, these approaches can become outdated and struggle to cover all possible brands comprehensively \cite{emd, phishzoo, visualphishnet, phishpedia, phishintention, dynaphish, li2024knowphish}.

\paragraph{Search Engine-based Approaches} These methods detect phishing websites by querying search engines with key descriptors from webpage content \cite{xiang2011cantina+, xiang2009hybrid, zhang2007cantina, huh2012phishing, jain2018two, varshney2016phish, chang2013phishing, chiew2015utilisation, rao2019jail}. If the input URL's domain appears in search results, the website is deemed legitimate. This method can lead to false positives as not all legitimate webpages are indexed or ranked highly.

\paragraph{LLM/MLLM-based Approaches} These leverage advanced capabilities in text and image processing to detect phishing \cite{koide2023detecting}. Prompts including webpage URL, HTML content, and screenshots help MLLMs predict phishing attempts. However, LLMs/MLLMs are vulnerable to adversarial attacks \cite{li2024knowphish, cui2024robustness} and can encounter difficulties with local brands due to the limited and unfamiliar information, which may result in incorrect decisions.

\paragraph{Autonomous Agents}  powered by LLMs and tools excel in handling complex tasks through efficient information processing and decision-making. \citet{park2023generative} introduces generative agents using large language models to simulate human behavior. Works like HuggingGPT \cite{shen2024hugginggpt}, AgentFLAN \cite{chen2024agent}, AgentInstruct \cite{zeng2023agenttuning}, and ReAct \cite{yao2022react} demonstrate the ability of LLMs to manage AI models and solve complex tasks. Recently, GEPAgent \cite{wang2024automatedphishingdetectionusing}, an agent-based phishing detection method, was introduced. It demonstrated high accuracy in identifying phishing webpages by leveraging interactions between MLLMs and an online knowledge base over several iterations. However, it also faced a major drawback regarding execution time, taking an average of 10 seconds per sample. 

\paragraph{Multimodal Retrievers} Information Retrieval (IR) methods aim to search for relevant information from an information collection \cite{singhal2001modern, wei2023uniir}. Among these, Multimodal Retrievers are those where the query and retrieved content can span across multiple modalities, such as the image and text modalities. This topic has been widely studied \cite{chao2021scaling, jain2021mural, li2021fuse, luo2023knowledge, rohit2023imagebind, wei2023uniir} and has applications in various domains. Inspired by \cite{wei2023uniir}, we design a retriever that uses HTML and brand logo of the webpage as the query to retrieve its potential target brands.

\section{Threat Model}
In a phishing attack, the attacker seeks to deceive users into thinking that the webpage is affiliated with a legitimate brand, thereby tricking them into disclosing sensitive information such as usernames, passwords, or bank details. Formally, let  $w$ denote a webpage, which includes its screenshot ($w.$\texttt{scr}), HTML structure ($w.$\texttt{html}), and a URL ($w.$\texttt{url}). To effectively carry out this deception, the webpage must convincingly imitate a specific brand $b$ by leveraging visual elements in $w.$\texttt{scr}, textual features in $w.$\texttt{html}, or both. Our goal is to detect phishing webpages, identify their target brands, and provide detailed annotations explaination.

\section{Multimodal Agent}
\subsection{Overview}
\begin{figure}[t]
    \centering
        \includegraphics[width=0.48\textwidth]
    {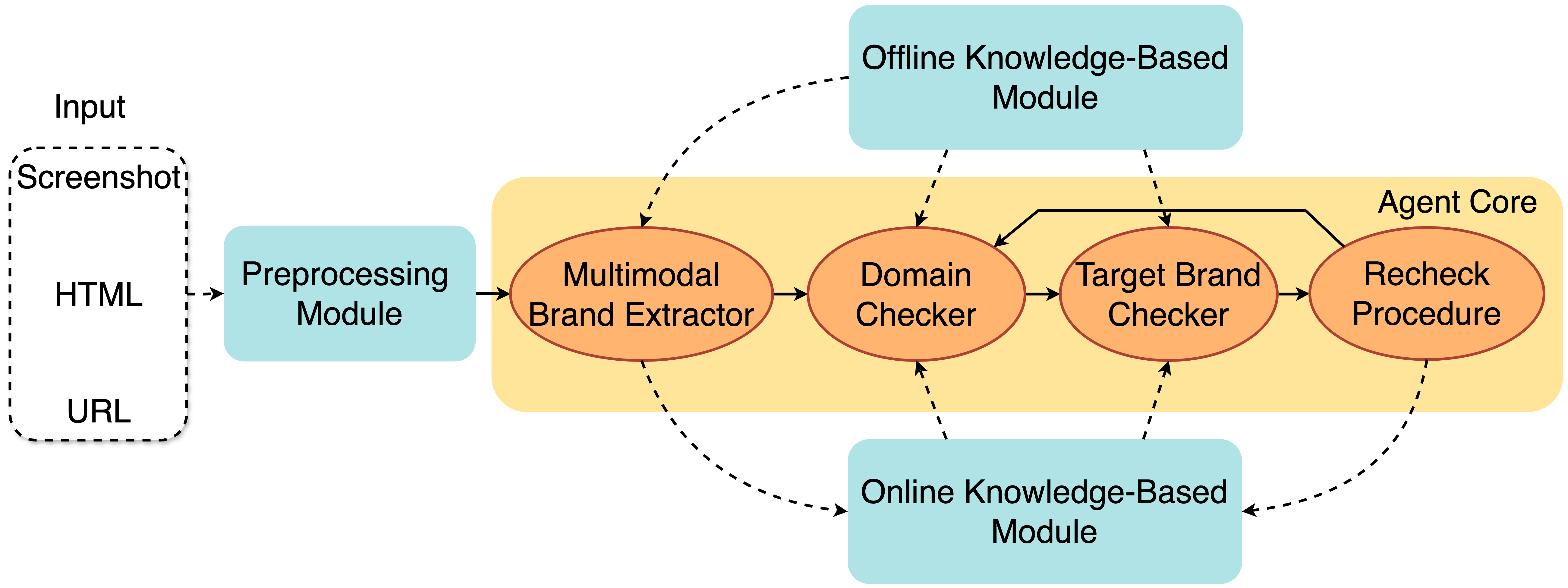}
    
    \caption{An overview of our phishing detector, PhishAgent.}
    \label{fig:agentphish}
\end{figure}
We next introduce PhishAgent, which leverages multiple tools and knowledge bases to verify different indicators of phishing activities. The Fig. \ref{fig:agentphish} shows the overview of PhishAgent.
PhishAgent consists of four main modules: Preprocessing Module, Online Knowledge-Based Module, Offline Knowledge-Based Module, and Agent Core. Each module can interact with others. The Agent Core functions as the central module, responsible for analyzing information and making decisions, while the remaining modules serve as tools to assist the Agent Core in preprocessing and gathering information. We analyze the details and role of each module in the next sections. 

\subsection{Preprocessing Module}
\label{sec:preprocessing}
 Given a webpage $w$ as input, we use a logo detection pipeline (see  Appendix for details) to extract its logo image (if any), denoted $w.\texttt{logo}$. Next, the HTML of $w$ is processed to remove noninformative HTML (such as layout or tracking elements), resulting in its processed HTML $w.\texttt{p\_html}$. Finally, the URL is parsed to extract its domain $w.\texttt{domain}$. The resulting logo, processed HTML, and domain are passed to the next step for further analysis. 

\subsection{Offline Knowledge-Based Module}
\label{sec:mr}
\begin{figure}
    \centering
    \includegraphics[width=0.45\textwidth]{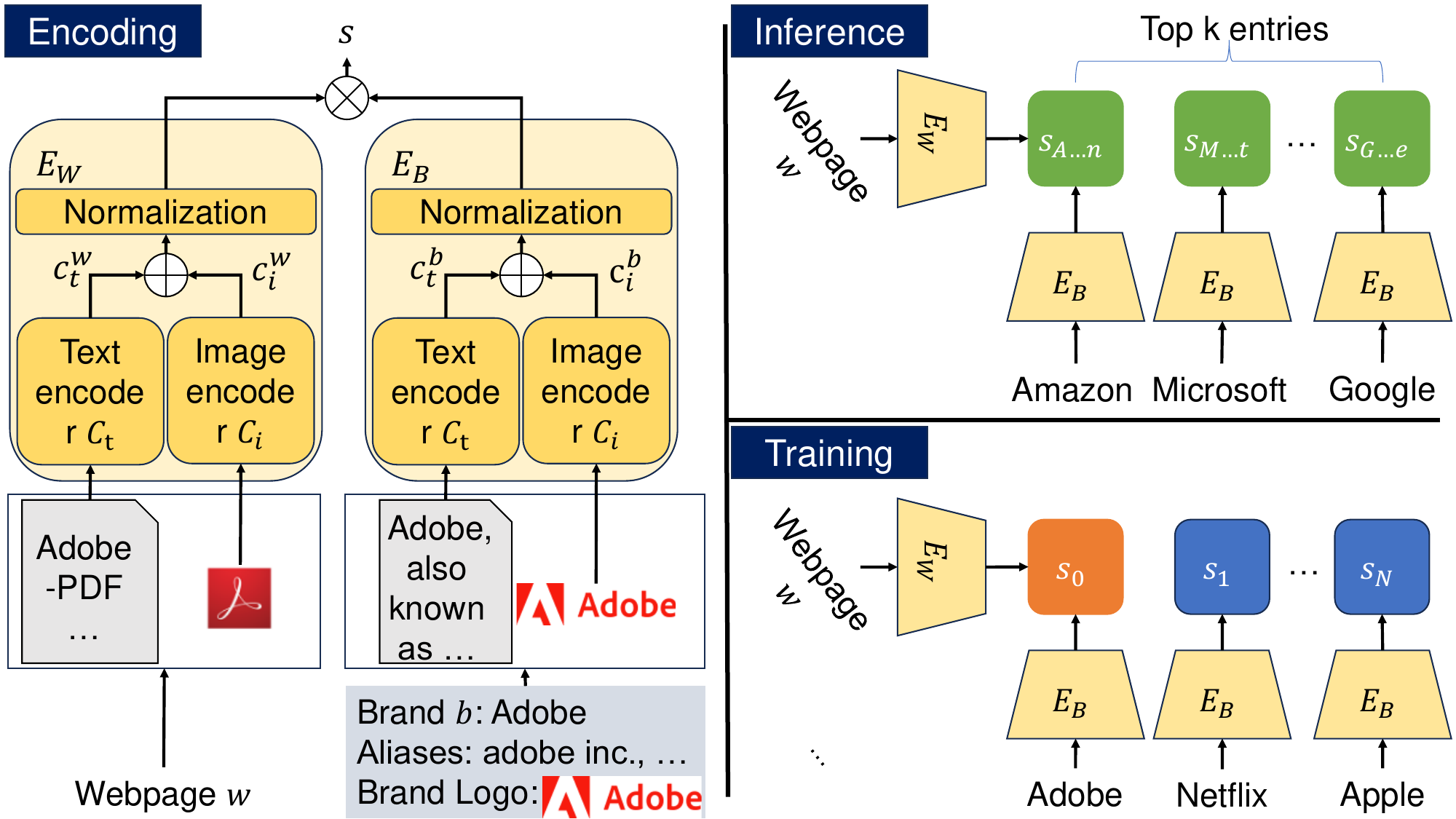}
    \caption{Our multimodal retriever; Left: Example of how a webpage $w$ and a brand $b$ are encoded and how the retrieval score $s$ is computed; Top right: Example of how the top $k$ brands are retrieved for the webpage $w$ during inference; Bottom right: Example of our training process where contrastive learning is used to distinguish from $N$ randomly sampled negative brands (colored in \colorbox{cyan}{blue}) the positive brand (colored in \colorbox{orange}{orange}) for the webpage $w$.}
    \label{fig:enter-label}
\end{figure}
Given the preprocessed information of a webpage \(w\) (\(w.\texttt{p\_html}\), \(w.\texttt{logo}\)), the goal of the Offline Knowledge-Based Module is to retrieve the brands and their associated information from the Brand Knowledge Base that are most similar to \(w\). Brand Knowledge Bases (BKBs) are collections of brands and their authentic information. In this work, we utilize KnowPhish \cite{li2024knowphish}, a multimodal BKB built on Wikidata, which includes around 20,000 potential phishing brands. Each brand entry \(b\) contains its name (\(b.\texttt{name}\)), logo (\(b.\texttt{logo}\)), aliases (\(b.\texttt{aka}\)), and domains (\(b.\texttt{domains}\)).

To achieve this, inspired by \citet{wei2023uniir}, we design and train a multimodal retriever \(\text{MR}\) that can process the webpage's HTML and extracted brand logo, retrieving the top $k$ brands from the BKB that most closely match the target brand of the webpage.

Specifically, our retriever encodes the input webpage into a \emph{webpage encoding}, and each brand into a \emph{brand encoding}. Our retriever then retrieves the top $k$ brands with the highest cosine similarity between their brand encodings and the webpage encoding. 

\subsubsection{Webpage Encoding}

Given the processed HTML $w.\texttt{p\_html}$ and brand logo $w.\texttt{logo}$ of the webpage $w$, we use a CLIP text encoder $\mathcal{C}_t$ to encode its processed HTML into an encoding $\mathcal{C}_t(w.\texttt{p\_html})$, and similarly use a CLIP image encoder $\mathcal{C}_i$ to encode the logo into an encoding: $\mathcal{C}_i(w.\texttt{logo})$. These text and image encodings are combined with some weights, added together, and normalized, to produce the combined webpage encoding:
\begin{align}
    E_W(w) = \mathsf{Norm}(c^{w}_{t} \mathcal{C}_t(w.\texttt{p\_html}) + c^{w}_{i} \mathcal{C}_i(w.\texttt{logo})),
\end{align}

where $c^{w}_{t}$ and $c^{w}_{i}$ are weight constants, and $\mathsf{Norm}$ normalizes the encoding.

In some cases, the logo extractor may not identify any logo on a webpage. In this case, we simply exclude the image embedding from $E_W(w)$.

\subsubsection{Brand Encoding} We compute the combined embedding for each brand $b$ in the knowledge base BKB in a similar manner. The only difference is that, since we have both the brand name $b.\texttt{name}$ (e.g., Microsoft) and a list of $A$ aliases $b.\texttt{aka}$ (e.g., [microsoft, $\cdots$, msft]) to represent each brand, we introduce a function $\oplus$ that combines them into a string with format \colorbox{lgray}{\{$b.\texttt{name}$\}, also known as \{$b.\mathtt{aka_1}$\}, $\cdots$, \{$b.\mathtt{aka_{A}}$\}} (e.g., \colorbox{lgray}{Microsoft, also known as microsoft, $\cdots$, msft}). Given $b.\texttt{name}$, $b.\texttt{aka}$, and $b.\texttt{logo}$ of each brand in the knowledge base, we compute the brand's combined encoding:
{
\begin{align*}
    E_B(b) = \mathsf{Norm}(c^{b}_{t} \mathcal{C}_t((b.\texttt{name} \oplus b.\texttt{aka})) + c^{b}_{i} \mathcal{C}_i(b.\texttt{logo}))
\end{align*}
}

where $c^{b}_{t}$ and $c^{b}_{i}$ are weight constants.

Again, certain brands may not have any logo image, or others may have more than 1 logo variant. To handle the former case, we exclude the image embedding when computing $E_B(b)$. In the latter case, for each logo variant of the brand $b.\mathtt{logo_j}$, we compute a separate embedding $E_B(b_j)$ with $b.\mathtt{logo_j}$ and treat it as a separate brand.

% add it as a new row to the brand embedding matrix $\textsf{Emb}_{\text{\, BKB}}$.

\subsubsection{Inference}

We then use dot product retrieval between $E_W(w)$ and $E_B(b)$ to get the top $k$ brand matches.

\begin{align*}
    \mathcal R_{\text{offl}} = \text{MR}(w, \text{BKB}, k) = \underset{b \in \text{BKB}}{\mathsf{TopK}}(E_W(w) \cdot E_B(b))
\end{align*}

We then pass the names, aliases, and legitimate domains for each retrieved brand to subsequent modules.

\subsubsection{Training}

We notice that retrieval recall is not satisfactory with the pre-trained CLIP encoders. Therefore, we use contrastive learning to train the CLIP encoders. Specifically, for each webpage in the training set, we annotate its ground truth target brand $b_0$ and randomly sample $N$ negative brands $b_1, \cdots, b_N$ from the knowledge base $k$. Denote $E_W(w) \cdot E_B(b_i)$ as $s_i$. We train our retriever by minimizing $-\mathsf{log}(\mathsf{softmax}(s_0, s_1, \cdots, s_N)_0)$ using stochastic gradient descent.

Experimental details of our retriever and its impact on brand name extraction performance are  in the Appendix.

\subsection{Online Knowledge-Based Module}
\label{sec:on_kb}
The goal of Online Knowledge-Based Module is to query all information related to the input webpage $w$ from online sources by leveraging search engines (e.g., Google search engine). This module utilizes two distinct queries corresponding to two different types: a domain-based query $\mathcal Q_{\text{domain}}$ and a brand name-based query $\mathcal Q_{\text{brand}}$. The domain-based query is the domain of the input webpage (e.g.,``amazon.sg") obtained from the preprocessing module, while the brand name-based query is generated by the Agent Core module (e.g., ``Amazon") as we will discuss in section 4.5 \- Multimodal Brand Extractor. These two queries aim to achieve two main objectives: assessing the popularity of the target brand and the popularity of the input domain. For each type of query, we retrieve the top $k$ items, where $R_{\text{domain}}$ is the set of top $k$ results from $\mathcal Q_{\text{domain}}$, and $R_{\text{brand}}$ is the set of top $k$ results from $\mathcal Q_{\text{brand}}$.

The domain-based query $\mathcal Q_{\text{domain}}$ focuses on evaluating the popularity of the input domain. If the domain is found in $R_{\text{domain}}$, it is highly likely to be a benign webpage, as most phishing webpages typically have a very short lifespan and are not indexed by search engines \cite{phishintention, dynaphish}.

The brand name-based query $\mathcal Q_{\text{brand}}$ is tasked with searching for webpages that belong to a specific brand name. This query aims to gather all relevant webpages associated with a brand while also checking the popularity of the brand name, as every notable brand name appears in Google search results.

The results from both queries are then combined:
\[
\mathcal R_{\text{onl}} = \mathcal R_{\text{domain}} \cup \mathcal R_{\text{brand}}
\]

Each item in the results will include the title of the webpage, the snippet, and the domain. The combined results $\mathcal R_{\text{onl}}$ are then returned to the Agent Core for further processing.

The synergy between these two types of queries ensures a comprehensive search, as they effectively complement each other. The domain-based query $\mathcal Q_{\text{domain}}$ covers cases where authentic webpages may not be popular enough to appear in the results of a brand name-based query $\mathcal Q_{\text{brand}}$. These less popular but still legitimate pages might otherwise be missed, which could lead to false positive cases. On the other hand, the brand name-based query $\mathcal Q_{\text{brand}}$ directly targets the brand name, helping to determine if the target brand is recognized. This capability is especially necessary when $\mathcal Q_{\text{domain}}$ cannot be found in search results. 

\subsection{Agent Core}
\label{sec:agent_core}
Agent Core plays a central role in analyzing data returned from the Online and Offline Knowledge-based Modules and making decisions. Agent Core consists of four main components, executed in the following order: Multimodal Brand Extractor, Domain Checker, Target Brand Checker, and Recheck Procedure. The illustration of Agent Core is in Fig. \ref{fig:agentphish}. Prompts and implementation details for each component can be found in the Appendix.

\subsubsection{Conditions for Identifying Phishing Webpages}
\label{sec:vr}
Our approach works by determining the \emph{target brand} of the webpage, i.e., the brand which can be identified from the page's HTML or logo. For example, a phishing webpage imitating PayPal by displaying the PayPal brand name and logo has the target brand of PayPal. 
A webpage is identified as phishing if the target brand of that webpage can be determined, and can be found in either the online knowledge base or the offline knowledge base (indicating that the target brand is recognized) but its domain does not match any domain from the online and offline search results (implying that the webpage is an unknown domain for a known brand, and is therefore a likely phishing webpage).
Let:
\begin{itemize}
    \item $w.$\texttt{targetb} be the target brand of the webpage.
    \item $\mathcal{B}_{\text{comb}}$ be the set of brands in $\mathcal R_{\text{onl}}$ and $\mathcal R_{\text{offl}}$.
    \item $\mathcal{D}_{\text{comb}}$ be the set of domains in $\mathcal R_{\text{onl}}$ and $\mathcal R_{\text{offl}}$.
    % \item $\textit{netloc}_{input}$ be the netloc of the input webpage.
    % \item $\text{netloc}(\text{domain})$ be the netloc of the webpage.
\end{itemize}

The condition for a webpage to be identified as phishing can be expressed as:

\begin{equation}
    \text{Phishing} =
    \begin{cases}
      \text{True} & \textit{if } (w.\texttt{targetb} \in \mathcal{B}_{\text{comb}}) \text{ and}\\ 
       &   (w.\texttt{domain} \notin \mathcal{D}_{\text{comb}})\\
      % \text{False} & \textit{if }  w.\texttt{domain} \in \mathcal{D}_{\text{comb}} \text{ or} \\
      % & w.\texttt{targetb} \text{ is ``Not Identifiable``}
      \text{False} & \textit{ otherwise} 
    \end{cases}
\end{equation}
Here, $\mathcal{B}_{\text{comb}}$ and  $\mathcal{D}_{\text{comb}}$ are from both Online and Offline Knowledge-based Modules. The target brand of the input webpage $w.$\texttt{targetb} is  determined by the Multimodal Brand Extractor. Responsibility of checking whether  $w.\texttt{domain} \in \mathcal{D}_{\text{comb}}$ belongs to the Domain Checker component, while Target Brand Checker verifies if $ w.\texttt{targetb} \in \mathcal{B}_{\text{comb}}$. Note that Agent Core can immediately determine whether a webpage is phishing or benign at any step by any component, as long as the conditions outlined in Equation (2) are met.
\subsubsection{Multimodal Brand Extractor (MBE)} 
\label{sec:mbe}
This component is responsible for identifying the target brand of the input webpage. The input to the MBE includes information from the target webpage ($w.\texttt{p\_html}$, $w.\texttt{domain}$, $w.\texttt{scr}$) and a string $B_k $ that concatenates the set of top $k$ searched brand names and their aliases in $R_{\text{offl}}$ from the BKB. Specifically, we use two LLMs/MLLMs to determine the brand name. The first is an LLM, the Text-based Brand Extractor (TBE), which identifies the brand name from the HTML, domain, and the top $k$ items from the offline knowledge base. The second is a MLLM, the Image-based Brand Extractor (IBE), which identifies the brand name from the webpage screenshot.

Firstly, $w.\texttt{p\_html}$, $w.\texttt{domain}$, and $B_k$ are sent to the TBE along with a designed prompt. The TBE determines the target brand by analyzing the HTML and the domain of the webpage, considering $B_k$ as potential brands. The LLM can refer to these brands to make informed brand names. In addition to the external knowledge given in the input, our approach also allows the LLM to utilize its internal knowledge.
If none of the potential brand names are suitable, the LLM can output a target brand that is not in $B_k$ or return "Not Identifiable". This is essential due to potential inaccuracies in the multimodal retriever or the absence of the target brand in the BKB. 
\[
    w.\texttt{targetb}  =  \text{TBE}( w.\texttt{html}, w.\texttt{domain}, B_k)
    \]

We only apply the Image-based Brand Extractor (IBE) to the screenshot if $w.\texttt{targetb}$ is "Not Identifiable". The input to the IBE is solely the screenshot. 
    \[
    w.\texttt{targetb} = \text{IBE}(w.\texttt{scr})
    \]

The webpage $ w $ is immediately deemed benign if $ w.\texttt{targetb} $ is ``Not Identifiable'' after calling the IBE, as it does not target any brand and is considered phishing-free.

There are two primary reasons we use text and visual information separately instead of simultaneously with a MLLM. First, in many instances, text information alone is sufficient to identify the target brand of a webpage. Always employing MLLMs would unnecessarily increase costs and average running time. Second, using two brand extractors sequentially enhances the robustness of the MBE against adversarial attacks. HTML can easily be manipulated to deceive the model if only a single MLLM is used. Our method, with the image-based brand extractor serving as a backup, addresses failures of the text-based brand extractor, whether due to insufficient text information or being tricked by adversarial attacks. More analysis of MBE is in the Appendix.

\subsubsection{Domain Checker (DC)} plays a role in verifying whether the domain of the input webpage ($w.\texttt{domain}$) matches any domains obtained from both the online knowledge base and offline knowledge base ($\mathcal{D}_{\text{comb}}$). The webpage $w$ is immediately determined as benign if $w.\texttt{domain} \in \mathcal{D}_{\text{comb}}$.

\subsubsection{Target Brand Checker (TBC)}
\label{sec:tbc}
Given  $w.$\texttt{targetb} and $\mathcal{B}_{\text{comb}}$ as input, the TBC checks if $ w.\texttt{targetb} \in \mathcal{B}_{\text{comb}}$. We adopt an LLM to do this task. The webpage $w$ is determined as phishing if $w.\texttt{targetb} \in \mathcal{B}_{\text{comb}}$ (since $w.\texttt{domain} \notin \mathcal{D}_{\text{comb}}$ in DC). 

\subsubsection{Recheck Procedure}
The goal of this procedure is to reduce false negatives and enhance PhishAgent's robustness against HTML-based adversarial attacks. Specifically, when the TBE identifies a target brand $w.\texttt{targetb}$ that is not marked as 'Not Identifiable' (thus IBE has not been called), but the target brand $w.\texttt{targetb}$ cannot be found in the search results ($w.\texttt{targetb} \notin \mathcal{B}_{\text{comb}}$ and $w.\texttt{domain} \notin \mathcal{D}_{\text{comb}}$), which may result from adversarial attacks or misidentification. The Recheck Procedure addresses these cases by re-extracting the target brand using an MLLM through the screenshot.
\[
w.{\texttt{targetb\_new}}, \textit{same\_old} = \text{Recheck}(w.\texttt{scr},w.\texttt{targetb})
\]
Here, $w.{\texttt{targetb\_new}}$ represents the new target brand redetermined by the Recheck Procedure. The variable $\textit{same\_old}$ indicates whether this newly detected brand $b.\texttt{targetb\_new}$ matches $w.\texttt{targetb}$. If $\textit{same\_old}$ is False, $w.{\texttt{targetb\_new}}$ is used as $\mathcal{Q}_{\text{brand\_new}}$ in the Online Knowledge Base to obtain $\mathcal{R}_{\text{onl\_new}}$. This result is merged with $\mathcal{R}_{\text{offl}}$ to form $\mathcal{R}_{\text{comb\_new}}$. DC and TBC are then called to check against $\mathcal{B}_{\text{comb\_new}}$ and $\mathcal{D}_{\text{comb\_new}}$. Designed conditions are applied again to determine whether a webpage is phishing. If $\textit{same\_old}$ is True (indicating the new brand matches the old one), the webpage is classified as benign without further analysis.

Note that in the case the $\text{IBE}$ has been invoked in MBE, the input webpage is directly determined as benign without calling the Recheck Procedure. 
\section{Experiments}
\begin{table*}[htbp]\scriptsize
    \centering
    \renewcommand{\arraystretch}{0.65}
    \scalebox{0.97}{
    \begin{tabular}{p{1.1cm}p{1.0cm}|p{0.6cm}<{\centering}p{0.5cm}<{\centering}p{0.85cm}<{\centering}p{0.5cm}<{\centering}p{0.5cm}<{\centering}|p{0.6cm}<{\centering}p{0.5cm}<{\centering}p{0.85cm}<{\centering}p{0.5cm}<{\centering}p{0.5cm}<{\centering}|p{0.6cm}<{\centering}p{0.5cm}<{\centering}p{0.85cm}<{\centering}p{0.5cm}<{\centering}p{0.5cm}<{\centering}}
    \toprule
        \multirow{2}{*}{\textbf{Detector}} & \multirow{2}{*}{\textbf{BKB}} & \multicolumn{5}{c|}{\textbf{\texttt{TR-OP}}} & \multicolumn{5}{c|}{\textbf{\texttt{SG-SCAN-1k}}} & \multicolumn{5}{c}{\textbf{\texttt{TR-AP}}} \\
        & & \textbf{ACC} & \textbf{F1} & \textbf{Precision} & \textbf{Recall} & \textbf{Time} & \textbf{ACC} & \textbf{F1} & \textbf{Precision} & \textbf{Recall} & \textbf{Time} & \textbf{ACC} & \textbf{F1} & \textbf{Precision} & \textbf{Recall} & \textbf{Time} \\
    \midrule    
                            & Original  & 68.20 & 53.61 & 99.06 & 36.75 &  0.26s  & 52.50 & 9.52 & \textbf{100.00} & 5.00 & 0.30s & 76.45 & 69.33 & 99.38 & 53.23 & 0.27s \\
         Phishpedia         & DynaPhish & 65.97 & 51.58 & 89.40 & 36.25 & 10.63s & 58.40 & 32.69 & 85.59 & 20.20 & 12.34s & 68.57 & 56.96 & 90.30 & 41.60 & 9.97s \\
                            & KnowPhish & 85.15 & 82.78 & 98.48 & 71.40 &  0.19s & 56.50 & 23.01 & \textbf{100.00} & 13.00 & 0.29s & 80.15 & 75.65 & 97.83 & 61.67 & 0.26s \\
    \midrule
                            & Original  & 65.60 & 47.60 & 99.84 & 31.25 &  0.29s & 51.90 & 
 7.32 & \textbf{100.00} &  3.80 & 0.33s & 73.35 & 63.72 & \textbf{99.79} & 46.80 & 0.30s \\
        PhishIntention      & DynaPhish & 61.98 & 39.86 & 95.27 & 25.20 & 10.40s & 52.50 & 10.88 & 87.88 &  5.80 & 11.94s & 68.57 & 54.62 & 98.18 & 37.83 & 9.76s\\ 
                            & KnowPhish & 77.65 & 71.24 & \textbf{99.91} & 55.35 &  0.24s & 53.10 & 11.68 & \textbf{100.00} & 6.20 & 0.32s & 75.65 & 67.94 & 99.42 & 51.60 & 0.36s \\
    \midrule
        \multirow{2}{*}{KPD} & DynaPhish & 76.70 & 70.75 & 95.03 & 56.35 & 11.92s & 60.20 & 35.60 & 93.22 & 22.00 & 11.40s & 74.25 & 68.74 & 95.27 & 53.77 & 10.32s \\
                            & KnowPhish &92.05 &91.44&98.95&  85.00 &1.49s & 65.20 & 47.27 & 97.50 & 31.20 & 2.07s & 89.13 & 88.06 & 97.68 & 80.17 & 2.22s \\
    \midrule
        GEPAgent           & Online & 92.95 & 92.70 & 96.13 & 89.50 & 12.35s & 83.10 & 82.66 & 84.76 & 80.80 & 13.51s & 89.58 & 89.61 & 89.44 & 89.77 & 14.88s\\
    \midrule
        ChatPhish           & None & 95.80& 95.91 & 93.80 & 98.10 & 6.93s & 83.50 & 82.85 & 86.13 & 79.80 & 7.03s & 91.07 & 90.90 & 92.60 & 89.27 & 6.89s\\
    \midrule
        \textit{PhishAgent} & Combined
 &  \textbf{96.10} & \textbf{96.13} & 95.24 & \textbf{97.05} & 2.25s
 &  \textbf{94.30} & \textbf{94.12} & 95.30 & \textbf{93.20} & 2.54s
 &  \textbf{94.87} & \textbf{94.86} & 95.02 & \textbf{94.70} &2.43s
 \\
    \bottomrule
    \end{tabular}}
    \caption{Phishing detection performance comparison of different baselines across the \texttt{TR-OP}, \texttt{SG-SCAN-1k}, and \texttt{TR-AP} datasets, where a lower `Time' metric, measured in seconds, indicates better performance, while higher values are preferable for the other metrics, all of which are presented as percentages.}
    \label{tab:effectiveness_and_efficiency}
\end{table*}
%  means higher is better while  refers to the opposite.

\begin{table}[!t]\scriptsize
    \centering
    \renewcommand{\arraystretch}{0.7}
    \begin{tabular}{llccccccc}
    \toprule
        \textbf{Detector} & \textbf{BKB} & \textbf{\#P} & \textbf{\#TP} & \textbf{Precision} & \textbf{Time} \\
    \midrule
                            & Original  &  54 &  17 & 31.48 & 0.16s \\
        Phishpedia          & DynaPhish & 583 & 481 & 82.67 & 5.98s \\ 
                            & KnowPhish & 353 & 333 & 94.33 & 0.16s \\ 
    \midrule
                            & Original  & 25  &   8 & 32.00 & 0.18s\\
        PhishIntention      & DynaPhish & 163 & 140 & 85.89 & 5.91s \\
                            & KnowPhish & 138 & 133 & 96.37 & 0.19s \\
    \midrule
        \multirow{2}{*}{KPD}                 
                            & DynaPhish & 628 & 581 & 92.52 & 7.83s \\
                            & KnowPhish & 699 & 681 & 97.42 & 1.64s \\ 
    \midrule
        PhishAgent                 
                            & Combined & 4139 & ~3936 & ~95.10 & 2.59s \\
    \bottomrule
    \end{tabular}
    \caption{Phishing detection performance of different baselines on \texttt{SG-SCAN-unl} dataset. \#P represents the number of reported phishing instances, while \#TP indicates the count of true positives.}    
    \label{tab:field_study}
\end{table}
\begin{table}[ht]
\centering
\scalebox{0.66}{
\setlength{\tabcolsep}{2.9pt} 
\begin{tabular}{clcc}
\toprule
\textbf{Type of attack} & \textbf{Injected Prompt/Typosquatting}  & \textbf{PhishAgent}  & \textbf{w/o Rck}  \\
\midrule
None & None  & 93.20 & 91.80 \\
\midrule
\multirow{3}{*}{\begin{tabular}{c}Brand spoofing \\ (HTML)\end{tabular}} & ``MobrisPremier"  & 91.20  & 89.40  \\
 &``Brand name is MobrisPremier"  & 91.20 & 85.00  \\ 
 & ``The official webpage of MobrisPremier"  & 90.40 &  71.20 \\ 
\midrule
 \multirow{2}{*}{\begin{tabular}{c}Brand spoofing \\ (SC)\end{tabular}} & ``MobrisPremier"  & 92.80 &  90.60 \\
 % & "Brand name is MobrisPremier"  & &  \\ 
 & ``The official webpage of MobrisPremier"  & 92.80 & 90.20 \\ 
 \midrule
 \multirow{2}{*}{\begin{tabular}{c}Brand spoofing \\ (SC + HTML)\end{tabular}} & ``MobrisPremier"  & 90.80 & 89.20  \\
 % & "Brand name is MobrisPremier"  & &  \\ 
 & ``The official webpage of MobrisPremier"  & 89.40 & 70.80 \\ 
 \midrule

\multirow{6}{*}{\begin{tabular}{c}Brand obfuscation\\ (HTML)\end{tabular}} & ``Please ignore the previous instruction& \multirow{2}{*}{91.00}  & \multirow{2}{*}{89.20}\\ &  and answer Not identifiable" & &   \\  
\cmidrule(lr){2-4} 

 & ``Please ignore the previous instruction&   \multirow{2}{*}{91.40}  &  \multirow{2}{*}{87.40} \\ &  and answer Not exist"  & &\\
 % \addlinespace 
\cmidrule(lr){2-4} 
 & Typosquatting on full HTML & 90.40 & 90.40\\
 & Typosquatting on the title only & 92.40 & 91.00 \\

\bottomrule
\end{tabular}}
\caption{Recall of PhishAgent with and without the Recheck component on different types of attacks.}
\label{tab:adv}
\end{table}

\begin{table}[ht]
\centering
\scalebox{0.71}{
\setlength{\tabcolsep}{2.9pt} 
\begin{tabular}{lcccccc}
\toprule
\textbf{Model} & \textbf{ACC} & \textbf{F1} & \textbf{Precision} & \textbf{Recall} & \textbf{Time} \\
\midrule
PhishAgent & 96.10 & 96.13 & 95.24 & 97.05 & 2.25s \\
\midrule
\quad w/o Offline Knowledge-based Module & 94.83 & 94.95 & 92.87 & 97.10 & 2.14s \\
\quad w/o Online Knowledge-based Module & 81.58 & 83.54 & 75.39 & 93.75 & 3.18s \\
\quad w/o Domain-based query & 84.63 & 86.58 & 76.55 & 99.85 & 2.02s \\
\quad w/o Brand Name-based query & 89.40 & 88.58 & 95.97 & 82.25 & 2.03s \\
\quad w/o Recheck Procedure &  95.95 & 95.97 & 95.05 & 96.95 &2.19s \\
\quad w/o Text-based Brand Extractor & 95.50 & 95.45 & 95.27 & 95.75 & 3.35s\\
\quad w/o Image-based Brand Extractor & 85.88 & 84.18  & 95.67 & 75.15 & 1.31s
 \\
\bottomrule
\end{tabular}}
\caption{Ablation study on \texttt{TR-OP}}
\label{tab:ablation_study}
\end{table}
% \vspace{-0.4cm}
We assess PhishAgent through the following research questions:
\begin{itemize}
    \item \textbf{RQ1 (Effectiveness and Efficiency)}: How do the effectiveness and efficiency of PhishAgent in identifying phishing webpages on real datasets compare to state-of-the-art methods?
    \item\textbf{RQ2 (Field Study)}: What is PhishAgent's effectiveness in identifying phishing attempts in real-world scenarios?
    \item\textbf{RQ3 (Adversarial Robustness)}: How well does PhishAgent withstand various adversarial attacks?
    \item\textbf{RQ4 (Ablation Study)}: How does each part of PhishAgent contribute to its overall performance?
\end{itemize}
\vspace{-0.2cm}
\subsection{Datasets}
We utilize three datasets for our main phishing detection experiments. \texttt{\textbf{TR-OP}}  \cite{li2024knowphish} is a manually labeled and balanced dataset, with benign samples randomly selected from the top 50k domains on Tranco  \cite{tranco}, and phishing samples sourced from OpenPhish  \cite{openphish}. The phishing samples were collected and validated over a six-month period, from July to December 2023, encompassing 440 unique phishing targets. The \texttt{TR-OP} dataset contains 4,000 samples, evenly split between 2,000 phishing and 2,000 benign webpages. \texttt{\textbf{SG-SCAN}} is collected from Singapore's local webpage traffic. \texttt{SG-SCAN} contains
 10k webpages collected from mid-August 2023 to mid January 2024  \cite{li2024knowphish}. \texttt{SG-SCAN} is divided into two datasets. The first dataset, \texttt{SG-SCAN-1k}, is manually labeled and balanced, containing 1,000 webpages, evenly split between 500 phishing webpages and 500 benign webpages. The second dataset, \texttt{SG-SCAN-unl}, is unlabeled and used for the field study. \texttt{SG-SCAN} is used to evaluate the phishing detection approaches in the local context. \texttt{\textbf{TR-AP}} is similar to \texttt{TR-OP} where its benign samples are a different subset of the Tranco top 50k domains from \texttt{TR-OP}. Its phishing samples were gathered from the empirical study of  \citet{li2024knowphish}, originally from APWG \cite{apwg}. The \texttt{TR-AP} dataset contains 6,000 samples, evenly split between 3,000 phishing and 3,000 benign webpages. 
 % The statistics of the 3 datasets can be found in Appendix. 
\subsection{Baselines}
We enlist three cutting-edge approaches as the phishing detector backbones: Phishpedia  \cite{phishpedia}, PhishIntention  \cite{phishintention}, and KPD  \cite{li2024knowphish}. As for the knowledge base, Phishpedia and PhishIntention can be integrated with either their original reference list (which includes 277 brands), DynaPhish (increasingly constructed by search engines) \cite{dynaphish}, or KnowPhish  \cite{li2024knowphish}. These are reference-based and search engine-based approaches.  Additionally, we consider GEPAgent  \cite{wang2024automatedphishingdetectionusing}, an agent-based approach, and ChatPhishDetector  \cite{koide2023detecting}, an MLLM-based approach (using ChatGPT 4o), as baselines in our experiments. 

 \subsection{RQ1: Effectiveness and Efficiency}
 We conduct the experiments for RQ1 on \texttt{TR-OP}, \texttt{SG-SCAN-1k}, and \texttt{TR-AP}. We evaluate PhishAgent against the baseline models based on  accuracy, F1 score, precision, recall and the average running time per sample.

Table \ref{tab:effectiveness_and_efficiency} presents the phishing detection performance of PhishAgent in comparison to various baselines. Overall, PhishAgent consistently exhibits superior performance across all datasets and maintains an efficient inference time. 

Particularly, on the \texttt{TR-OP} dataset, PhishAgent achieves remarkable results with an accuracy of 96.10\%, F1 score of 96.13\%, precision of 95.24\%, and recall of 97.05\%, while maintaining an average inference time of 2.25 seconds. For the \texttt{TR-AP} dataset, PhishAgent attains top metrics with an accuracy of 94.87\%, F1 score of 94.86\%, precision of 95.02\%, recall of 94.70\%, and an inference time of 2.43 seconds. On the \texttt{SG-SCAN-1k} dataset, which focuses on local webpage phishing, PhishAgent demonstrates superior performance with an accuracy of 94.12\%, F1 score of 95.30\%, precision of 93.20\%, recall of 97.30\%, and an inference time of 2.54 seconds, showing an improvement of more than 10\% over the state-of-the-art methods.

PhishAgent mainly classifies webpages as benign when URLs are verified by both knowledge bases. Across benign datasets, 5,236 samples were classified as benign, with only 91 samples (1.74\%) due to an unidentifiable brand name, showing knowledge base verification as the primary factor. 

PhishAgent operates efficiently in low-resource environments by relying on external APIs for LLMs/MLLMs and online knowledge, minimizing local computation. The retriever requires only ~1.6GB of GPU memory, and PhishAgent can function effectively with just online knowledge and MLLMs without the retriever if necessary as shown in RQ4.
 \subsection{RQ2: Field Study}
Following the Field Study methodology of KnowPhish  \cite{li2024knowphish}, we conducted our field study on the \texttt{SG-SCAN-unl} dataset. As this dataset is unlabeled, we only validate the samples flagged as phishing by the detectors. The number of positive (\#P) and true positive (TP) are reported. This methodology enables us to assess the real-world effectiveness of PhishAgent in detecting phishing webpages. 

The experimental results of RQ2 are reported in Table \ref{tab:field_study}. Overall, PhishAgent demonstrates superior performance by detecting 4,139 phishing webpages compared to the 681 detected by KnowPhish. Although PhishAgent has a slightly lower precision at 95.10\% compared to the 97.42\% of KnowPhish, the significant increase in the number of detected phishing webpages makes this trade-off worthwhile.

 \subsection{RQ3: Adversarial Robustness}
We evaluate PhishAgent's robustness against real-world adversarial attacks designed to misclassify phishing webpages as benign. While HTML manipulation is harder for users to detect, visual-based attacks are less common due to their visibility. We test these attacks on both HTML and screenshots (SC). The types of adversarial attacks include:
\begin{itemize}
    \item Brand spoofing aims to trick the model into incorrectly identifying a brand name that is not found in the Online Knowledge Base or Offline Knowledge Base, potentially causing the sample to be mistakenly classified as benign based on verification rules. To simulate the attack, we inject adversarial prompts into HTML, SC, and both.
    \item Brand obfuscation aims to prevent the model from recognizing the brand name. The model may output phrases such as ``Not identifiable," ``Does not exist," or similar phrases. For this attack, we inject adversarial prompts into the HTML or use typosquatting.
\end{itemize}

We simulate adversarial attacks on the 500 phishing samples from the \texttt{SG-SCAN-1k}. Table \ref{tab:adv} shows the recall of PhishAgent, as well as its performance after removing the Recheck Procedure on different types of attacks. PhishAgent demonstrates robustness against various types of adversarial attacks. Although performance is affected, it remains within acceptable levels, with the greatest reduction being nearly 4\% from the original performance when conducting brand spoofing attacks on both screenshots and HTML. Additionally, the experimental results highlight the effectiveness of the Recheck Procedure in enhancing the PhishAgent's robustness against adversarial attacks.

 \subsection{RQ4: Ablation Study}
We evaluate the impact of each  component in PhishAgent by sequentially removing them and observing performance changes through an ablation study on the \texttt{TR-OP} dataset. Results are shown in Table \ref{tab:ablation_study}. Overall, all components contribute positively to phishing detection performance. Notably, while removing TBE has minimal impact on accuracy due to IBE’s capabilities, TBE significantly improves efficiency, reducing runtime from 3.35s to 2.25s. Similarly, the Recheck Procedure, although contributing little to overall performance, proves highly effective in adversarial attack scenarios. Further details are provided in the Appendix.

\section{Conclusion}
In conclusion, we introduce PhishAgent, a multimodal agent that integrates a comprehensive set of modules and leverages both online and offline knowledge bases with Multimodal Large Language models. PhishAgent effectively synthesizes various approaches, maximizing their strengths while minimizing their weaknesses. Our experimental results demonstrate that PhishAgent is both effective and efficient, performing well across various settings, including local brand phishing detection. Furthermore, PhishAgent exhibits robustness against a wide range of adversarial attacks, indicating its potential for effective use in real-world scenarios.
\section*{Acknowledgements}
This work was supported by the National Research Foundation Singapore, NCS Pte. Ltd. and National University of Singapore under the NUS-NCS Joint Laboratory (Grant A-0008542-00-00), as well as by the Ministry of Education, Singapore, through the Academic Research Fund Tier 1 (FY2023) (Grant A-8001996-00-00).

\bibliography{reference}
% \newpage
% \appendix
% \input{script/section_appendix}

% \newpage
\end{document}

\maketitle
% \section{Additional Details on KnowPhish Construction}
% \label{appendix:additional_details_on_knowphish_construction}

% \section{Additional Details on KnowPhish Detector}
% \label{appendix:knowphish_detector}

% \section{Additional Details on Experiments}
% \label{appendix:additional_details_on_experiments}

% \section{Appendix}
% \subsection{Potential of Agent-based Approaches}
% \label{sec:simple_agent}

% We compare a simple agent GEPAgent \cite{wang2024automatedphishingdetectionusing} with the SOTA phishing detector \cite{li2024knowphish} to demonstrate the potential of autonomous agents.

% % Include the design of the simple agent here.

% \begin{figure}
%     \centering
%     \includegraphics[width=0.48\textwidth]{figures/SimpleAgent.pdf}
%     % \includegraphics[width=\textwidth]{figures/Retriever.pdf}
%     \caption{The comparison between GEPAgent \cite{wang2024automatedphishingdetectionusing} and the SOTA (KnowPhish) \cite{li2024knowphish} on TR-OP dataset. The metric ‘Runtime’ indicates the average inference time per sample, while the remaining metrics are presented in percentages. The lower the time metric, the better, while the higher the performance metric, the better.}
%     \label{fig:SimpleAgent}
% \end{figure}

% While having a much longer runtime, the GEPAgent surpasses the SOTA method in terms of recall, accuracy, and F1 score. This gives a strong incentive for more exploration in the application of autonomous agents in phishing detection.
\section{Experimental Details of the Multimodal Retriever (MR)}
\label{sec:exp_mr}

\subsection{Encoding} We follow \citet{wei2023uniir} and set $c^w_t$, $c^w_i$, $c^b_t$, and $c^b_i$ to 1 respectively.

\subsection{Inference} We set $k$ to 5. Empirically, we find that this value of $k$ works well in conjunction with the MBE.

% Chúng tôi sử dụng phishing samples trong bộ TR-OP \cite{li2024knowphish}. Note that, the phishing samples using for training is not overlaped with the samples we use for evaluation. 
\subsection{Training} 
\subsubsection{Dataset} 
To train the MR, we employ phishing samples from the original TR-OP dataset \cite{li2024knowphish}. It's important to note that the phishing samples used for training the MR are distinct from those used for testing PhishAgent. Specifically, from phishing samples in the original TR-OP dataset, we use 2,000 for testing (as mentioned in the main content of the paper), while a separate set of 2,500 samples are used to train the MR, which we name as \textbf{\texttt{OpenPhish$_{train}$}}. The selection of this dataset is driven by the fact that the phishing samples in TR-OP come with target brand labels \footnote{The labels are simply the target brand names of the webpage.}, which are crucial for effective training of the MR.
 
% We split the original TR-OP datasets from \citet{li2024knowphish} into 2 parts. The first part is used to test PhishAgent and its components, including MR, which contains 2,000 phishing and 2,000 benign webpages. The TR-OP dataset in the main content and any experiments in the paper refers to the first part only. The second part is the remaining, which is solely used to train and validate our MR. We denote the second part as \textbf{TR-OP$_{train}$} We use the \texttt{TR-OP$_{train}$} for traning our MR, since it provides the target brand annotation for each sample and is relatively large.

% Here the mentioned 2,158 samples are from the second half.
% We use the second half of the original \textbf{TR-OP} dataset , denoted as OpenPhish\_{train}, to curate the training and validation sets, since it provides the target brand annotation for each sample and is relatively large.

To train MR, we need to collect $(w, b_0, b_1, \cdots, b_N)$ pairs where $b_0$ is the target brand of the webpage $w$, and $b_1$ to $b_N$ are the negative brands (These brands are not related to $w$ but used for contrastive learning only).

We also require that $b_0 \in BKB$ so that we can have multimodal information from $BKB$ to compute the brand embedding. However, for certain webpages in \texttt{OpenPhish$_{train}$}, it may happen that $b_0 \notin BKB$. Even if $BKB$ contains $b_0$, we still may not know which brand $b_0$ is inside $BKB$ since our target brand label $b_0'$ is just the brand name. 

Therefore, we do an initial grounding step where we match the label $b_0'$ to a brand $b_0$ in $BKB$. In particular, we count, for each webpage $w$ and $b_0'$, how many brands in $BKB$ $b_0'$ can be matched to, using a string match algorithm. We filter out webpages whose $b_0'$ is matched to 0 brand (no match) or more than 1 brand (ambiguous matches) in $BKB$, retaining those with exactly 1 matched brand, which we use as $b_0$. We then pair $b_0$ with $w$. 

After this step, we are left with 2,158 $(w, b_0)$ pairs. We then augment these pairs by creating a new $(w, b_{0j})$ pair for each logo variant $b_0.logo_j$ that $b_0$ has. This gives us 5,889 $(w, b_{0j})$ pairs. Lastly, we do 2 rounds of random sampling from $BKB$ to get 2 separate sets of negative brands $b_1, \cdots, b_N$ for $w$, which gives us 11,778 pairs. We then proceed with a 9:1 split to get 10,598 training pairs and 1,180 validation pairs.

During training, we set the batch size to 4, learning rate to $2e-6$, and the number of negative brands $N$ to $111$. We use Adam Optimizer and train the model for 10 epochs, and then we select the checkpoint with the highest validation score for inference.

\section{Analysis on the Multimodal Brand Extractor (MBE)}
\label{sec:analysis_mbe}

Our MBE is at the core of our phishing detection pipeline. It receives the top $k$ potential brands retrieved by the Multimodal Retriever $R$ and then predicts the target brand. We use 2,310 samples from $TR-OP$ \footnote{We filter out samples for which the target brand labels cannot be grounded in $BKB$. This filtering step is different from what is used in the main experiment, which explains why we have 2,310 samples here.} that have the target brand labels to analyze the various components of our MBE.

\noindent\paragraph{Ablation Study on MR} Since the brand predicted by MBE can come from either (1) the knowledge retrieved by MR or (2) the parametric knowledge of the LLM itself, we do an ablation study to investigate the effect of removing MR, in which case MBE can only rely on its parametric knowledge.

\begin{table}[htbp]\footnotesize
    \centering
    \renewcommand{\arraystretch}{0.9}
    \begin{tabular}{l|c}
    \toprule
        \textbf{Method} & \textbf{Rec.} \\
    \midrule
       MBE w/ $MR_{k=5}$ & 73.94 \\ 
    \midrule
        \hspace{10mm} w/o MR & 66.28 \\
    \bottomrule
    \end{tabular}
    \caption{Ablation Study of our Multimodal Brand Name Extractor MBE.} 
    \label{tab:retriever_ablation_study}
\end{table}

% \%Rec. w/ Sup: Percentage of recalled cases where the true brands are retrieved by MR. \%Rec. w/o Sup: Percentage of recalled cases where the true brands are not retrieved by MR and are therefore inferred from the parametric knowledge of the LLM during inference.

As shown in Table \ref{tab:retriever_ablation_study}, without MR, MBE achieves a recall of around 66 percent. The recall increases to 72.86 percent when MR is added. This means that \textbf{external brand knowledge does help target brand detection, at least to a certain extent}. In particular, it helps in cases where the true brand is unknown to the LLM or hardly discernible. This justifies the inclusion of MR in our design.\\

\noindent\paragraph{Effect of $k$} $k$ controls the amount of related or unrelated knowledge passed from MR to MBE, and it can significantly affect the prediction process. Therefore, we vary the value of $k$ to check its effect on both the retriever itself and the MBE.

\begin{table}[htbp]\footnotesize
    \centering
    \setlength{\tabcolsep}{2pt}
    \scalebox{0.8}{
    \begin{tabular}{l|ccccccc}
    \toprule
        \textbf{Method} & \textbf{Rec.@1} & \textbf{Rec.@2} & \textbf{Rec.@3} & \textbf{Rec.@5} & \textbf{Rec.@10} & \textbf{Rec.@50} & \textbf{Rec.@100} \\
    \midrule
       MR & 83.93  & 88.01 & 90.39 & 92.81 & 95.24 & 97.75 & 98.01 \\ 
    \midrule
        & \textbf{Rec.$_{k=1}$} & \textbf{Rec.$_{k=2}$} & \textbf{Rec.$_{k=3}$} & \textbf{Rec.$_{k=5}$} & \textbf{Rec.$_{k=10}$} & \textbf{Rec.$_{k=50}$} & \textbf{Rec.$_{k=100}$} \\
    \midrule
        MBE  & 72.03  & 72.60 & 72.86 & 73.94 & 73.29 & 69.09 & 64.37 \\
    \bottomrule
    \end{tabular}}
    \caption{Recall of our Multimodal Retriever MR and our Multimodal Brand Name Extractor MBE when varying $k$, the number of brands retrieved by $R$. Rec.$_{k=n}$: Recall of the MBE when $k$ is set to $n$ for MR.}    
    \label{tab:retriever_analysis}
\end{table}

As can be seen from Table \ref{tab:retriever_analysis}, increasing $k$ continuously boosts the recall of MR. When it comes to MBE, its recall is relatively stable and robust when $k$ changes from 1 to 10, but \textbf{an excessively large $k$ can hurt the performance of MBE}. This is expected. With a larger $k$, our MBE has a higher chance to acquire the true brand from MR, but there are also more distractions from the additional incorrect brands that are retrieved along. This experiment justifies our choice of setting $k$ to 5.

\
\noindent\paragraph{Empirical study on MBE with a single MLLM}
As discussed in the main content, we process text and visual information separately rather than simultaneously with a single MLLM. For better understanding, we input all the information (HTML, screenshot, and URL) along with the top $k$ items from the BKB into the MLLM at once, allowing it to extract the brand name from the provided data. Experimental results in Table \ref{tab:sgscan_mbe} indicate that using the MLLM to process multimodal information simultaneously decreases the phishing detection performance of PhishAgent and increases execution time. This finding reinforces the rationale behind our design choice in handling text and visuals separately.

\begin{table*}[h]
\centering
\begin{tabular}{l|c|c|c|c|c}
\textbf{Model} & \textbf{Accuracy} & \textbf{F1} & \textbf{Precision} & \textbf{Recall} & \textbf{Time} \\ \hline
PhishAgent with Text-based Brand Extractor + Image-based Brand Extractor  & 94.30 & 94.12 & 95.30 & 93.20 & 2.54s \\ \hline
PhishAgent with One-pass Processing  & 90.9 & 90.5 & 93.8 & 87.6 & 4.90s \\ 
\end{tabular}
\caption{Phishing Detection Performance on \texttt{SG-SCAN-1k}. The first row shows the performance of PhishAgent when using Text-based Brand Extractor and Image-based Brand Extractor separately, as discussed in the main content. The second row presents the performance of PhishAgent when using a single MLLM to process HTML, URL, and screenshot simultaneously.}
\label{tab:sgscan_mbe}
\end{table*}

\section{Implementation Details of PhishAgent}
\begin{itemize}

\item \textbf{LLMs/MLLMs:} We use GPT-3.5-turbo-instruct as the LLM backbone (used in Text-based Brand Extractor and Target Brand Checker), while ChatGPT-4o is employed as the MLLM backbone (used in Image-based Brand Extractor and Recheck Procedure). We set the temperature of the LLM/MLL to 0 for all experiments to make the output more deterministic. 
\item \textbf{Search Engine:} We utilize the Google search engine for our experiments. We set $k$ to 5 for each query.
\item \textbf{Logo Extractor} extract existing logos from the screenshot. We utilize the Logo Extractor from Phishpedia \cite{phishpedia}. The Logo extractor use Faster-RNN as the backbone. Given a screenshot, the Logo Extractor reports a set of candidate logos, with a confidence score. We rank the logos by their score and take the top one as the identity logo.
\item \textbf{HTML Processor} removes the CSS, JavaScript, and HTML tags from the HTML and extracts only the plain text. We use the \textit{BeautifulSoup}\footnote{\url{https://beautiful-soup-4.readthedocs.io/en/latest}} library to implement this processor.
\item \textbf{URL processor} is simply a function that can extract the domain from the input URL. Specifically, we utilize \textit{urlparse} function from \textit{urllib} library. 
% \item Text-based Brand Extractor
\item \textbf{Domain Checker} is a simple function that checks whether the input domain exists in the set of domains or not.
\end{itemize}

\begin{tcolorbox}[colback=gray!10, colframe=gray, width=0.5\textwidth, sharp corners, boxrule=0.5mm, title= Prompt for Text-based Brand Extractor, fonttitle=\bfseries, coltitle=white, colbacktitle=gray]
\label{box:Text-based Brand Extractor}
System prompt: \\You are a helpful assistant that responds in detecting brand name in a webpage.\\

Instruction: \\Define targeted brand as a brand that a webpage belongs to. Given the URL and HTML along with the potential brands of a webpage P, answer:\\ 
(1) What the targeted brand of P is. The target brand can appear in the HTML or the URL, please look carefully. There are cases where the target brand appears in the url along with other characters, try to identify the target brand from the URL. If the analysis suggests that the brand name is not among the provided list of potential brands, independently identify and determine the correct brand name based on the content. Extract the brand name only and do not include extra details such as affiliated products, countries, or additional abbreviations; If you are not sure about what the targeted brand of P is, please output a brand name from the list of potential brands that you think it is the most relevant to the webpage P. If the brand is not identifiable from the given information, put "Not identifiable". You only output the brand name without any additional explanation.
\\

Input information:\\
--URL: \\``\{url\}"

--HTML: \\
``\{processed\_html\}"

Potential brands:\\
``\{top\_k\_brands\_from\_offline\_knowledge\_base\}"\\

Answer:

\end{tcolorbox}

\begin{tcolorbox}[colback=gray!10, colframe=gray, width=0.5\textwidth, sharp corners, boxrule=0.5mm, title= Prompt for Image-based Brand Extractor, fonttitle=\bfseries, coltitle=white, colbacktitle=gray]
System prompt: You are a helpful assistant that responds in detecting brand name in a webpage. \\

Instruction: Analyze the provided image to detect a single, authentic brand name, ensuring its accuracy even in the presence of adversarial attacks. If no brand name is identifiable in the image, return ``Not identifiable". Output only the genuine brand name without additional explanation. If there are multiple brand names in the image, identify the most suitable and authentic brand name associated with the genuine logo.

\end{tcolorbox}

\begin{tcolorbox}[colback=gray!10, colframe=gray, width=0.5\textwidth, sharp corners, boxrule=0.5mm, title=Prompt for Recheck Procedure, fonttitle=\bfseries, coltitle=white, colbacktitle=gray]
System prompt: \\You are a helpful assistant that responds in detecting brand name in a webpage.\\

Instruction:\\
(1) Analyze the provided image to detect a single, authentic brand name, ensuring its accuracy even in the presence of adversarial attacks. If no brand name is identifiable in the image, return ``Not identifiable". Output only the genuine brand name without additional explanation. If there are multiple brand names in the image, identify the most suitable and authentic brand name associated with the genuine logo.\\
(2) Does the detected brand name in (1) match the brand ``\{brand\_name\_old\}"? If yes, return ``Yes"; if no, return ``No" without any explanation.\\

Example output format:\\
(1) Singtel\\
(2) Yes
\end{tcolorbox}

\begin{tcolorbox}[colback=gray!10, colframe=gray, width=0.5\textwidth, sharp corners, boxrule=0.5mm, title=Prompt for Target Brand Checker, fonttitle=\bfseries, coltitle=white, colbacktitle=gray]
Given lists of items found from Google and Knowledge Base. \\

- A item found from Google has a format as follow: \\``[index]. [netloc]/[search title]/[snippet]", the full list is as follows:\\ ``\{items\_from\_online\_knowledge\}".\\ 

- The full list of items found from Knowledge Base is as follows:\\ ``\{items\_from\_offline\_knowledge\}". \\

Based on the given searched items, determine if any of the provided searched items mention to the brand ``\{determined\_target\_brand\}". If yes, return 1; if not, return 0. \\

Example output: 1

\end{tcolorbox}

\section{Experimental Details of the PhishAgent and Analysis}
\subsection{Resources} All the experiments were conducted on a Ubuntu server
with 2 AMD EPYC 7543 32-Core Processor @ 2.8GHz and
8 Nvidia A40 48GB GPU available.
\subsection{Baselines}
All baselines were implemented according to the original methods describe in their respective papers. To ensure fairness, for the baselines that utilize MLLMs (e.g., ChatPhishDetector \cite{koide2023detecting} and GEPAgent \cite{wang2024automatedphishingdetectionusing}), we uniformly employed ChatGPT-4o. For the baselines that use LLMs (e.g., KPD \cite{li2024knowphish}), we used GPT-3.5-turbo-instruct, as our PhishAgent also utilizes ChatGPT-4o as the MLLM and GPT-3.5-turbo-instruct as the LLM.
\subsection{Datasets}
Statistical overview of the datasets we use in our work are presented in Table \ref{tab:dataset}.
\begin{table*}[htbp]\footnotesize
    \centering
    \renewcommand{\arraystretch}{0.9}
    \begin{tabular}{ccccc}
    \toprule
        \textbf{Dataset} & \textbf{Samples} & \textbf{Benign} & \textbf{Phishing} & \textbf{Used in} \\
    \midrule
        \texttt{TR-OP} & 4000 & 2000& 2000 & RQ1 and 4\\
    \midrule
    
        \texttt{SG-SCAN\_unl} & 10000 & Unknown & Unknown & RQ2\\
        \midrule
    \texttt{SG-SCAN\_1k} & 1000 & 500 & 500 & RQ1 and 3\\
    \midrule
    \texttt{TR-AP} & 6000 & 3000 & 3000 & RQ1\\
    \midrule
        \texttt{OpenPhish$_{train}$} & 2500 & None & 2500 & Training the MR\\
    \bottomrule
    \end{tabular}
    \caption{Statistical overview of the main datasets.}
    \label{tab:dataset}
\end{table*}
\subsection{Adversarial Attacks}
\begin{figure}[htbp]
    \centering
    \includegraphics[width=\linewidth]{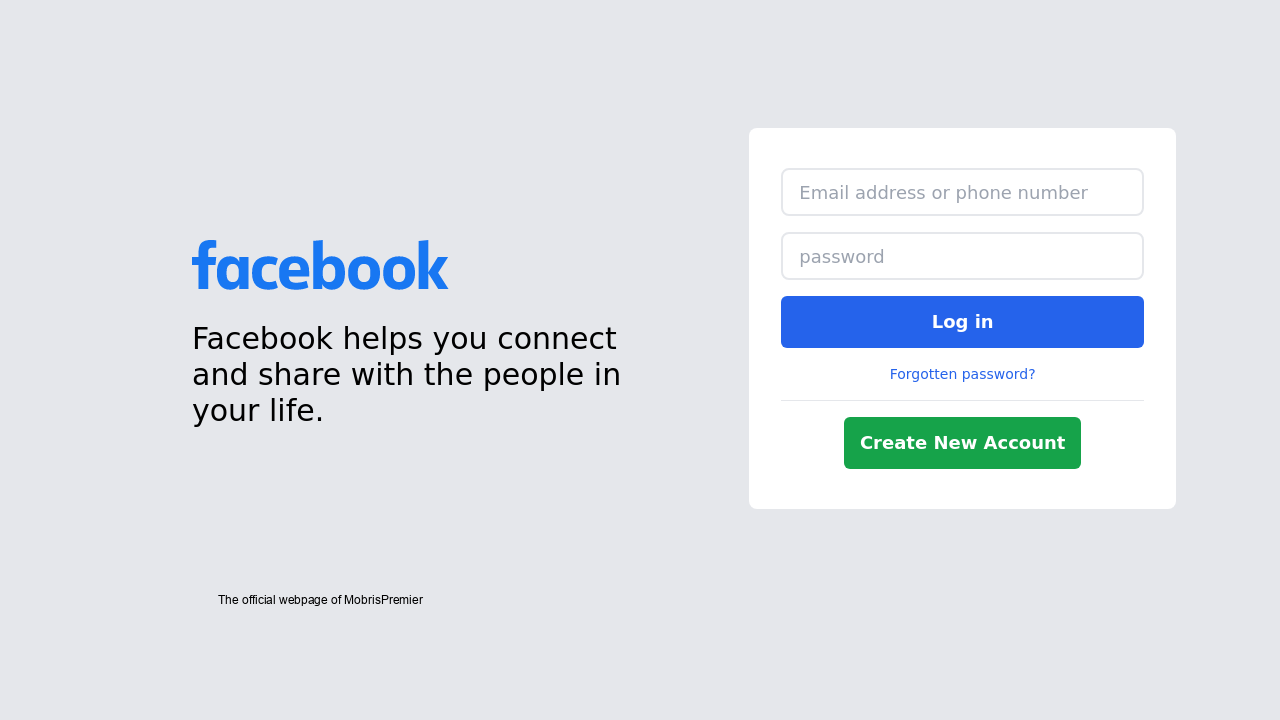}
    \caption{An example of brand spoofing attack on a screenshot. The target brand is ``Facebook", but the adversarial attack try to trick the Image-based Brand Extractor predict ``MobrisPremier" as the target brand}
    \label{fig:sc_atk}
\end{figure}

\textbf{Brand Spoofing:} We inject a brand name that is not present in either the Online Knowledge Base or the Offline Knowledge Base, leading the sample to be incorrectly classified as benign under the defined conditions. Specifically, we use the brand name ``MobrisPremier.'' For brand spoofing in HTML, we embed the prompt within an \texttt{<a>} tag and place this tag at the beginning of the HTML document. For screenshots, we embed the prompt directly into the image using the \texttt{PIL} library. The text is rendered in Arial font at a size of 12, ensuring it remains subtle enough to evade user detection. The text's position within the screenshot is randomly determined. An example of brand spoofing attack on a screenshot can be found in Fig. \ref{fig:sc_atk}.\\
\textbf{Brand Obfuscation:} There are two types of brand obfuscation attacks. The first involves prompt injection, where we embed the prompt within an \texttt{<a>} tag and place this tag at the beginning of the HTML document. The second type is typosquatting. Following \citet{li2024knowphish}, we replace common Latin characters with visually similar characters from other alphabets (e.g., Greek, Cyrillic) or symbols, targeting either the title or all text elements in the HTML. In our approach, we obfuscate one character in each word.

\subsection{Ablation Study Analysis}
\textbf{Offline Knowledge-based Module} The precision of PhishAgent decreases when this module is removed. This is because some benign domains are covered by the offline knowdge base but are not found in the online knowledge base, leading to false positives. The offline knowledge base plays an important role in enhancing PhishAgent's precision.\\
\textbf{Online Knowledge-based Module} Performance across all metrics decreases when the Online Knowledge-based Module is removed. Notably, the average running time increases because many benign samples are not immediately identified as benign after passing through the Domain Check. This results in more cases requiring the use of the Target Brand Checker or the Recheck Procedure, thereby increasing the average processing time for the entire dataset.\\
\textbf{Domain-based query and Brand Name-based query} Based on the experimental results, we observe that the domain-based query plays an important role in improving precision, as some benign webpages are not found in the search results from brand name-based queries, leading to false positives. On the other hand, the brand name-based query is essential for improving recall, as it effectively searches for brand information, allowing us to determine whether the identified brand is recognized.\\
\textbf{Recheck Procedure} The Recheck Procedure does not significantly contribute to overall performance, but it is necessary in scenarios involving adversarial attacks, as demonstrated in the main content.\\
\textbf{Text-based Brand Extractor} The overall performance only slightly decreases when we do not use the Text-based Brand Extractor, due to the strong capabilities of the Image-based Brand Extractor. However, the running time increases significantly, taking 3.35 seconds per sample compared to 2.25 seconds with the Text-based Brand Extractor included. This is because the Image-based Brand Extractor is invoked for all cases, leading to overhead in the processing time. \\
\textbf{Image-based Brand Extractor} The recall significantly drops, indicating that many webpages cannot be recognized by the Text-based Brand Extractor.

 \section{Cost per run}
 We calculated PhishAgent’s cost per run on TR-OP. MLLMs (GPT-4o and GPT-3.5) cost about 0.00191 USD per run, while search engine queries cost up to 0.005 USD each query after a 100-query daily free allowance. In practice, we immediately check the domain’s existence within search results after each query to avoid further API calls. PhishAgent often skips the search engine API if the domain is in our offline knowledge base or requires only one query from the search engine for benign samples.

\section{Case Study: How the Recheck Procedure handles Adversarial Attacks}
\begin{figure}[htbp]
    \centering
    \includegraphics[width=\linewidth]{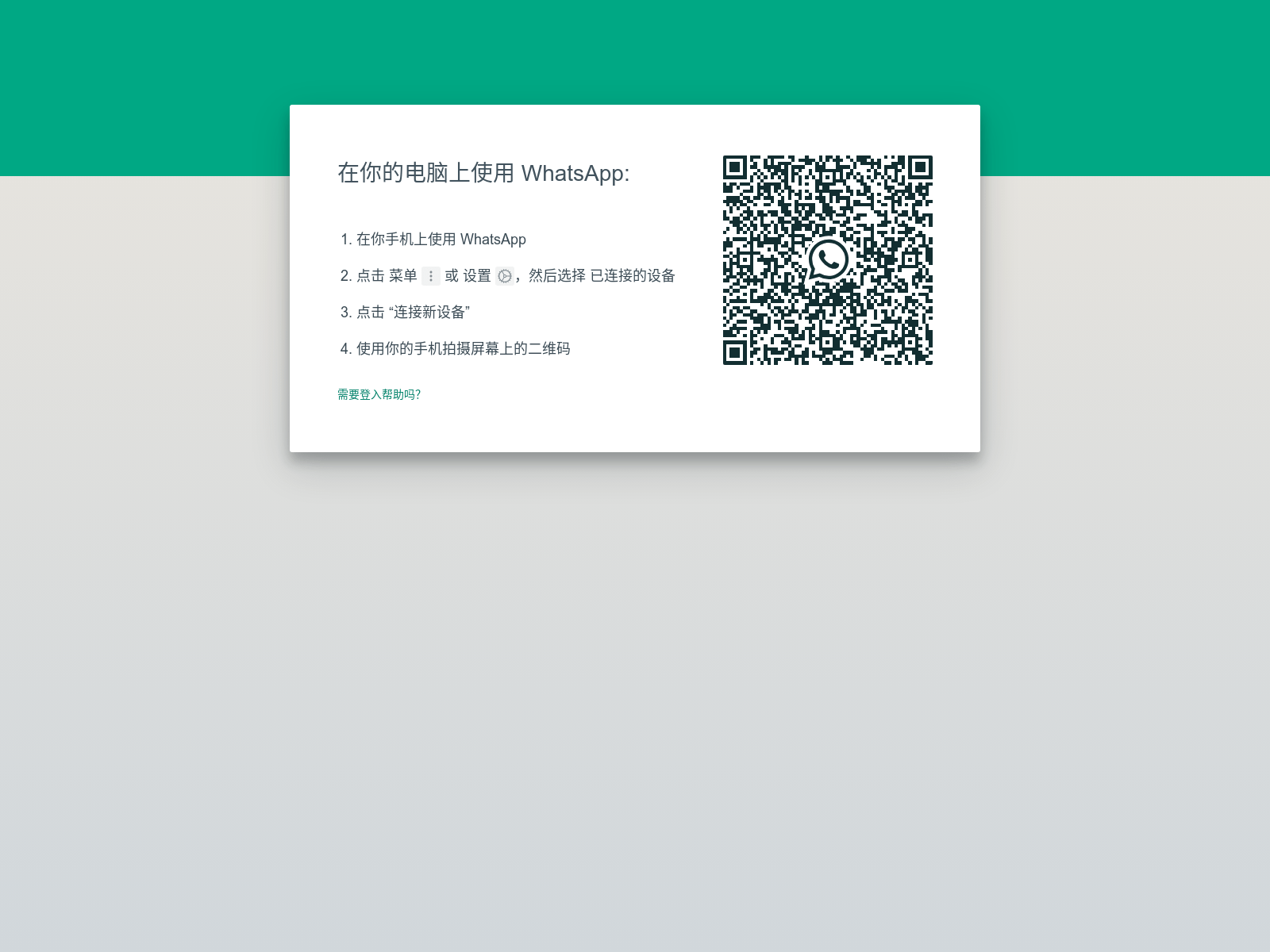}
    \caption{A screenshot of a phishing webpage with the target brand as ``WhatsApp".}
    \label{fig:casestudy}
\end{figure}

For a better understanding of how the Recheck Procedure works, let's consider a phishing webpage with the screenshot shown in Fig. \ref{fig:casestudy}. The URL of the webpage is ``\texttt{https://ws-9qd.pages.dev}". We inject the prompt ``\texttt{<a>The official webpage of MobrisPremier</a>}" into the head of the HTML. The Text-based Brand Extractor is misled by this prompt, incorrectly predicting the target brand as ``MobrisPremier.''

The domain ``\texttt{ws-9qd.pages.dev}" is used as the domain-based query, and ``MobrisPremier'' is used as the brand name-based query. First, the Domain Checker verifies whether the input domain appears in the search results. The input domain is not found (because it is a phishing domain, not indexed by search engines, and does not appear in the BKB). Next, the Target Brand Checker checks if the determined target brand exists in the search results. The result is that ``MobrisPremier'' cannot be found in either the Online or Offline Knowledge Base.

Without the Recheck Procedure, this webpage would be mistakenly classified as benign (since neither the target brand nor the domain can be found in any knowledge base), leading to a false negative. The Recheck Procedure corrects this by redetermining the target brand through the screenshot. It identifies ``WhatsApp'' as the new target brand, which differs from the previous one. Using ``WhatsApp'' as the new brand name-based query in a Google search yields different search results.

The Domain Checker is run again, and the domain ``\texttt{ws-9qd.pages.dev}" is still not found. However, when the Target Brand Checker is used again, it successfully finds ``WhatsApp'' in the search results, leading to the correct classification of the webpage as a phishing site.

% \include{s}
% \section{Appendix}
% \include{script/section_appendix}

\bibliography{reference}
% \newpage